\documentclass[12pt,preprint]{aastex}   
\def\paren#1{\left( #1 \right)}
\def\brace#1{\left\{ #1 \right\}}
\def\bra#1{\left[ #1 \right]}

\def\Mesz{M\'esz\'aros~}
\begin{document}
\title{Bulk Compton Emission in the GRB Internal Shock Model}
\author{Ryo Takagi\altaffilmark{1} and Shiho Kobayashi\altaffilmark{2,3}}
\altaffiltext{1}{Department of Physics, Tokyo Institute of Technology,
                 Meguro, Tokyo 152-8551, Japan}
\altaffiltext{2}{Center for Gravitational Wave Physics,
                 Pennsylvania State University, University
		 Park, PA 16802}
\altaffiltext{3}{Department of Astronomy and Astrophysics, 
                 Pennsylvania State University, University
		 Park, PA 16802}
\begin{abstract}
 We investigate the role of the bulk Compton-scattering process in the
 internal shock model. We numerically show that the radiation field
 created by internal shocks does not affect the efficiency of the energy
 conversion in the shocks even if fireball outflows have highly
 nonuniform velocity distributions. However, the bulk Compton-scattering 
 produces an additional high energy component especially when
 variability in the outflows arises because of a
 modulation of the mass injected into a constant energy flow. We show
 that an isotropic energy of $10^{50-51}$ ergs could be radiated in the
 range of 100MeV-GeV through the scattering process, provided that
 the shell's Lorentz factor varies between 10 and $10^4$. 
 \end{abstract}
\section{Introduction}
Gamma-ray bursts (GRBs) and afterglows are well described by the
relativistic fireball model, in which an explosive flow of relativistic
matter is released from a compact central engine (e.g. a massive stellar
collapse or a compact stellar merger). The kinetic energy of the flow is 
converted into internal energy by shocks, and it is radiated via
synchrotron process. These shocks can be due to collision with the
ambient medium (external shocks) or shocks inside the flow itself due to
nonuniform velocity (internal shocks). The highly variable temporal
structure observed in most of bursts indicates that internal shocks are
responsible for the production of GRBs. As internal shocks do not
dissipate all the kinetic energy, the remaining energy is dissipated
later by external shocks and produces an afterglow. 

Since the energy radiated in the afterglow phase is comparable or even
smaller than that in the prompt gamma-ray phase (Kumar \& Piran 2000;
Freedman \& Waxman 2001; Lloyd-Ronning \& Zhang 2004), 
GRB outflows should be highly irregular. If we represent the irregular
flow by a succession of relativistic shells with random Lorentz factors
(Kobayashi, Piran \& Sari 1997; Daigne \& Mochkovitch 
1998; Guetta, Spada \& Waxman 2001), the distribution of the Lorentz
factors should have a large dispersion to achieve a high conversion 
efficiency in the internal shocks (Beloborodov 2000; 
Kobayashi \& Sari 2001).

Recently, Lazzati, Ghisellini and Celotti (1999) pointed out that, if 
the Lorentz factor distribution is very broad, the radiation drag
process could affect the conversion efficiency and spectrum of the
internal shock emission (see also Gruzinov \& \Mesz 2000; Ghisellini
2003; Ghisellini et al. 2000). When the Lorentz factors of shells are 
distributed in a large
interval, fast shells would move in the radiation field produced by the
previous collisions, and thus Compton-scatter these photons
(bulk Compton-scattering). The fast
shells could be decelerated before colliding into other shells. This is
of significant interest, because high energy $\gamma$-rays could be 
produced as a consequence of the internal shock model. Observations 
of high energy radiation will provide valuable information about GRB 
outflows and their central engines.  In this Letter, we study the bulk
Compton process by using a numerical random shell model.
We discuss the implication to the Gamma-Ray Large Area Space Telescope
(GLAST).

\section{Bulk Compton-Scattering}
In the internal shock model, central engines are assumed to produce many
relativistic expanding shells with a random distribution of Lorentz
factors. A collision of two shells is the elementary process to
dissipate the kinetic energy of the shells (e.g. Kobayashi et al. 1997). A
rapid shell catches up with a slower one and the two merge to form a single
one. Electrons are heated by shocks, and the internal energy  is then
radiated via synchrotron emission.

Consider such a merged shell with a Lorentz factor $\gamma$, emitting an
internal energy $E$. When a fast shell with a Lorentz factor 
$\Gamma \gg \gamma$ collides into the merged shell, the fast shell 
scatters the photons emitted from the merged shell. In the comoving
frame of the merged shell, the internal energy $E^\prime \sim E/\gamma$
is emitted isotropically. Then, the fast shell is moving with the relative 
Lorentz factor $\Gamma^\prime \sim \Gamma/2\gamma$ into the photon
field, scattering a fraction $\sim \min(1,\tau)$ of these photons to energies
higher by a factor of $\sim 2\Gamma^{\prime 2}$ where $\tau$ is the optical
depth of the fast shell. The factor of $\sim 2$ in the boost factor takes
into account that the distribution of the photons which the fast
shell scatters is not isotropic, and that head-on scatterings are
favored (Ghisellini et al. 2000).
Using the mass of the fast shell $M$, 
we can estimate the energy extracted from the fast shell as 
\begin{equation}
Mc^2d\Gamma\sim - \frac{\min(1, \tau)}{2}  \paren{\frac{\Gamma}{\gamma}}^2 dE.
\label{eq:eom}
\end{equation}
The fast shell should have a smaller Lorentz factor $\Gamma_{dec}$ when
it catches up to the merged shell. It is given by
\begin{equation}
\Gamma_{dec} \sim  \Gamma 
\paren{1+\frac{\min(1,\tau)\Gamma E}{2\gamma^2Mc^2}}^{-1}.
\end{equation}
A fraction of the kinetic energy $\sim Mc^2 (\Gamma-\Gamma_{dec})$ is 
radiated away through the scattering process.

In the standard fireball model, the optical depth $\tau$ of a fireball
is close to unity $\tau \lesssim 1$ around the internal shock radius, because 
the parameters are chosen so as to make $\tau<1$ enabling a non-thermal
spectrum, and the ``preferred'' model uses larger baryon loads
(e.g. Gruzinov \& \Mesz 2000). Therefore, if merged shells radiate a
significant fraction of their kinetic energies through the internal 
shock synchrotron process, as it is assumed in the standard model, a 
large velocity contrast in the flow $\Gamma \gg \gamma$ could lead to an
efficient production of high energy $\gamma$-rays.   
\section{Numerical Shell Model}
We perform Monte Carlo simulations, and discuss how the bulk 
Compton-scattering affects the evolution of multiple shells. We
estimate the conversion efficiency 
from the kinetic energy of shells to high energy radiation 
in the bulk Compton-scattering.
\subsection{Outline of the Model}
We consider a flow consisting of $N$ shells. Each shell is 
characterized by four variables: a Lorentz factor $\gamma_i$, a mass
$m_i$, a width $l_i$ and a position $R_i$. We assume that the initial
Lorentz factor of each shell is distributed uniformly in logarithmic
space between $\log \gamma_{min}$ and  $\log \gamma_{max}$. The initial
masses are assumed to be correlated with the Lorentz factors as $m
\propto \gamma^s$.  For $s =-1$ and $0$, the shells initially have
equal energy and equal mass respectively, and for $s =1$ the shells
initially have equal density under an assumption of the equal shell
width. We assume a constant value $l$ for the initial widths and the
initial separations between the shells. Then, the initial positions of
the shells are $R_i=2(i-1)l$.   

The evolution of the shell system in time is basically the same as in
Kobayashi et al (1997), but the radiation drag effect is taken into
account. Because of the relativistic beaming effect, the radiation from
a merged  shell with a Lorentz factor $\gamma$ is concentrated in the
forward direction. A large fraction of photons are emitted around
$\theta \sim 1/\gamma$, and they propagate to the radial direction with
a velocity of $\sim c \cdot \cos(1/\gamma)$ or equivalently an effective
Lorentz factor of $\sim \gamma$. 
We regard the radiation from each collision as a photon packet
with the Lorentz factor of the merged shell $\gamma$. Since merged
shells will further collide with other shells, the Lorentz factors of 
the mergers would significantly change during the evolution. The Lorentz
factors of photon packets are assumed to be constant after the
productions.  

We turn now to describe the evolution of the multiple shell system. 
There will be numerous collisions between the shells. We denote by
the index $j$ the $j$-th collision which takes place at a time
$t_{j}$. Given the velocities of the shells $\beta _{i}$ and their
separations at $t_{j-1}$ we calculate for all pairs $(i,i+1)$ the
collision times. We then find the minimal collision  time $\delta t_{j}$.  
After rearranging shells from earlier positions $R_{i}(t_{j-1})$ to 
$R_{i}(t_{j})\equiv R_{i}(t_{j-1})+c\beta _{i}\delta t_{j}$, we consider 
the radiation drag effect. Each photon packet is characterized by a
position $r_n$, an energy $E_n$ and a Lorentz factor $\eta_n$
$(n=1,..,j-1)$. To estimate the radiation drag effect on 
shell $i$, we find all the packets $\brace{k}$ which the $i$ shell
overtakes between $t_{j-1}$ and $t_j$. The Lorentz factor of the $i$ 
shell at $t_j$ is evaluated by 
\begin{equation}
\gamma_i(t_j) =  \gamma_i(t_{j-1}) \paren{1+\min(1,\tau_i)
\sum_k \frac{\gamma_i(t_{j-1})E_k}{2\eta_k^2m_ic^2}}^{-1},
\end{equation}
where $\tau_i$ is the optical depth of the $i$ shell at $t_j$, and we take a
sum over the subset $\brace{k}$. The difference of the kinetic energies
$m_i c^2 \bra{\gamma_i(t_{j-1})- \gamma_i(t_j)}$ is emitted from the $i$
shell as high energy photons. Since a fraction of the photons in the
relevant packets $\brace{k}$ are scattered by the $i$ shell, we reduce the
energy as $E_k(t_j)=\bra{1-\min(1,\tau_i)}E_k(t_{j-1})$. 

We estimate the internal energy produced by the $j$-th collision
by using the energy-momentum conservation (eq. (2) in Kobayashi et
al. 1997). We assign $1/3$ of the internal energy to the $j$-th photon
packet produced in this collision if the optical depth of the merged
shell is less than unity. We assume that the remaining internal energy
is transformed to the kinetic energy of the merged shell. If the optical
depth is larger than unity, the whole internal energy is converted to
the kinetic energy of the merger. We find the next collision
and proceed until there are no more collisions, i.e. until all
the shells have merged to form a single shell or until the shells are
ordered with increasing value of the Lorentz factors (the outermost 
and innermost are fastest and slowest, respectively). 

Photon packets can boost shells in principle when they catch up with
slow shells. However, the boost effect is of the order 
of $\tau dE$ instead of $(\Gamma/\gamma)^2 \tau dE$ which is the
order of the radiation drag effect (see  eq. (\ref{eq:eom})). We find
that the boost effect is negligible in the evolution of the multiple 
shell system. In this letter, we discuss only the radiation drag effect.  

\subsection{Efficiency}
For $N=300$ shells, $\gamma_{min}=10$,  
$l/c=1$ sec and the initial kinetic energy  
$E_{kin,iso}=\sum m_i c^2\gamma_i =10^{53}$ erg, 
we have evaluated the conversion efficiencies of internal shocks and 
the bulk Compton-scattering. The mean efficiencies and their
standard deviations for 100 realizations are listed in Table 1. In the
third column, ``No Scattering'' indicates the internal shock 
efficiencies obtained without the scattering process. 
We can see that the efficiencies 
of the internal shocks are not influenced by the radiation drag effect 
significantly even when the velocity dispersion is large 
$\gamma_{max}/\gamma_{min} \gg 1$. 

While the internal shock process is most efficient in the equal mass
cases, the bulk Compton-scattering is most efficient in the equal energy 
cases. Recently Nakar and Piran (2002) studied time scales in GRB
temporal profiles. Their results imply that the variability in
wind's Lorentz factors arises because of a modulation of the mass
injected into a  constant energy flow (equal energy shells). If this 
is the case, a significant amount of energy, corresponding to
$1-10 \%$ of the prompt gamma-ray energy (GRB energy), might be 
radiated as high energy $\gamma$-rays through the bulk Compton-scattering. 

A large number of collisions  $\sim N$ happen during the evolution of
the multiple shells. The first collisions occur between the slowest
shells $\sim \gamma_{min}$ and fast shells $\gg \gamma_{min}$ around $R
\sim \gamma_{min}^2 l$. In equal energy cases, the Lorentz factors of  
the merged shells produced by the first collisions are comparable to 
that of the slowest shells $\sim \gamma_{min}$ (eq. (1) in Kobayashi et
al. 1997). This property allows other fast shells to catch up with 
the merged shells at small radii $\sim 2\gamma_{min}^2 l$. Since the fast 
 shells still have larger optical depth $\propto R^{-2}$ at the 
small collision radii, and the relative Lorentz factors are larger, 
the photons emitted from the first mergers are efficiently scattered
into high energies.
In equal density cases, the Lorentz factors of the first mergers are
close to $\sim \gamma_{max}$. It is difficult for other shells to catch 
up to the mergers. Since such collisions happen at large radii
and the relative velocities are small, the bulk Compton-scattering is
expected to be inefficient. In equal mass cases, the Lorentz 
factors of the first mergers take intermediate values 
$\sim (\gamma_{min}\gamma_{max})^{1/2}$. The efficiency of the 
scattering process is also expected to be intermediate.

We classify shell collisions occurring in the optically thin region 
into four cases: (I) Collisions happen between two shells, both of which
have never experienced collisions. Since we are interested in the
radiation drag effect, we count only collisions in the optically thin 
region. Even if a shell has collided with another shell in the optically 
thick region, we regard the merger as an unexperienced shell.
(II) While the outer slow shell went through collisions 
at least once, this is the first collision for the inner 
rapid shell. (III) The outer slow shell has never interact with 
other shells, and the inner rapid shell has experienced collisions 
at least once. (IV) The other cases. Assuming the same parameters as in
Table 1 and $\gamma_{max}=10^4$, we have performed numerical simulations
for the three mass distributions, and we have analyzed all shell
collisions in the optically thin region.  
We summarize the results for 100 realizations in Table 2. The radiation
drag effect is expected to be most significant in type II
collisions. Actually, in the equal energy case for which the bulk
Compton-scattering is most efficient, about $50\%$ of collisions are
classified into the type II. On the other hand, the fraction of type III
collisions is larger than that of the type II in the equal density
case. These results are consistent with the above discussion. 

Efficient bulk Compton-scattering requires the optical depth of 
shells to be close to unity $\tau \lesssim 1$ at internal 
shock radii. In fig 1, we plot efficiencies as functions of 
the variability time scale $l/c$, and we discuss how the optical 
depth is sensitive to the choice of the 
variability time scale. If it is very short $l/c\sim 1$ms,
most collisions happen at small radii. 
Since the shells still have large optical depth, the internal shock
emission is highly suppressed, and consequently only a small amount 
of energy is up-scatted to high energies. As a larger value of $l/c$ is 
assumed, a larger fraction of collisions happen in the optically thin
region. Around $l/c\sim$ a few sec, all the collisions happen in the 
optically thin region, and the internal shock efficiency (thin solid line) 
is saturated. The efficiency of the bulk Compton-scattering (thick solid line) 
also increases as a larger $l/c$ is assumed, because the
internal shocks create thicker photon field. However, if the variability
time is very long $\gtrsim 10$s, the optical depth of the shells at
collision radii is too small to up-scatter the photon field
effectively. For our typical parameters, the efficiency of the bulk
Compton-scattering peaks around $l/c \sim 0.1-1$ sec, which are
comparable to the typical variability time scales observed in GRB
profiles. 

All the numerical results obtained so far are for 
$E_{kin,iso}=10^{53}$ ergs, which is the 
average value of what we derive from observations. The 
distribution of isotropic GRB energies spans at least two orders of 
magnitude. Evaluating the efficiencies for $10^{52}$ ergs (dashed lines
in fig 1) and for $10^{54}$ ergs (dashed-dotted lines), we found that
the asymptotic values of internal shock 
efficiencies $\sim 11\%$ and the peak values of scattering 
efficiencies $\sim 3\%$ are the same, and that the efficiency curves 
just shift to left or right.
As we can expect from the optical depth of shells 
$\tau \propto E_{kin,iso}/l^2$, the saturation time scale for $10^{54}$ergs,
at which the internal shock efficiency is saturated and the scattering 
efficiency peaks, is one order larger than that for
$10^{52}$ergs. For the same distribution of Lorentz factors, 
the efficiency of internal shocks does not depend on 
$E_{kin,iso}$ if all collisions happen in the optical thin
region. The deceleration effect $d\Gamma$ by radiation field is larger
for a smaller baryon loading (a smaller $E_{kin,iso}$). However, the energy
extracted from shells depends on the mass of shells (baryon loading)
also, and it is determined by the photon field energy as
$Mc^2d\Gamma \propto dE$. The peak values of the scattering 
efficiencies should have similar values, because the internal shock
efficiencies are the same at the saturation time scales.

We have assumed that $1/3$ of the internal energy produced by 
each collision goes into radiation. Assuming a smaller value
of $0.1$ or $0.01$, we re-evaluated the efficiencies in
the equal energy case of $\gamma_{max}=10^4$ in table 1.
The internal shock efficiencies are $3.6\pm0.2 \%$ and $0.37\pm0.02 \%$ 
for the conversion factors of 0.1 and 0.01, respectively, while
the scattering efficiencies are respectively $1.1 \pm 0.3\%$ and
$0.14\pm0.04\%$. If less photons are produced, less photons would be 
scattered. The ratio of internal shock and scattering efficiencies is
almost constant.   

We found that the internal shock process is more efficient than
the bulk Compton-scattering. The main reason is that internal shocks
occur at a rather wide range of radii. Since the optical depth of shells
rapidly decreases as $\tau \propto R^{-2}$, the Compton-scattering
process  is effective only at small radii at which only a small
fraction of the kinetic energy of the flow is converted to photon field. We
have analyzed collisions occurring in an equal energy case with 
the same parameters as in fig 1 and $l/c \sim 1.5$s (the scattering
process is most efficient for this $l$), and we obtained the following
results. The internal shock 
emission starts around $R\sim 4 \times 10^{13}$cm, and the shocks at
$R<$ a few $10^{15}$cm contribute effectively to the total internal
shock radiation of $\sim 10^{52}$ergs. $10\%$, $50\%$ and $80\%$ of the
radiation are emitted within $R \sim 10^{14}$ cm ,
$5\times10^{14}$cm and $2\times10^{15}$cm, respectively. The minimum
internal shock radius $\sim 4\times 10^{13}$cm is consistent with the
photosphere radius $R_{ph}$ at which mergers becomes optically thin. It
is given by $R_{ph}\sim (\sigma_T E_{kin,iso}/4\pi N \gamma_{min}m_p
c^2)^{1/2}$ $\sim 3.4\times 10^{13}
(E_{kin,iso}/10^{53}$ergs$)^{1/2}(N/300)^{-1/2}
(\gamma_{min}/10)^{-1/2}$ cm. The bulk Compton-scattering process starts
around $\sim 2R_{ph}$, at which the optical depth of scatter shells with  
$\Gamma \gg \gamma_{min}$ is already much smaller than unity, because their
masses are smaller than the masses of the mergers. Consequently, the
Comptonization parameter $y=\tau\times$(relative Lorentz factor)$^2$ 
is on the order of unity at $ R \lesssim 10^{14}$cm and it decreases
rapidly at larger radii. $10\%$, $50\%$ and $80\%$ of the bulk Compton
emission are produced within $R\sim 9\times 10^{13}$cm, $2\times
10^{14}$ cm and $4\times 10^{14}$cm, respectively. Although the
scattering process is very effective $y\sim 1$ at small radii $R \lesssim
10^{14}$cm, the energy of the photon field created by the
internal shocks at the small radii is $\sim 10\%$ of the total
internal shock radiation, and then the efficiency from the kinetic
energy to high energy radiation in the bulk Compton-scattering is 
of the order of $1\%$. 
If we assume even larger contrasts of Lorentz factors
($\gamma_{max}=10^5$ or larger), scattering shells have larger 
Comptonization parameters $y \gg 1$ at the first scattering radius 
$\sim 2R_{ph}$. However, the scattering emission with large 
relative Lorentz factors is suppressed by the Klein-Nishina effect.
The resulting efficiency of the scattering process is about 4-5 $\%$ 
for $\gamma_{max}=10^{5-6}$.

The spectra of the high energy radiation produced by the bulk
Compton-scattering are shown in fig 2 for two example cases. Assuming
that all the primary photons produced by shell collisions have the
typical GRB frequency of $100$ keV, we have integrated up-scattered
photons to construct the spectra. Bursts of high energy emission
are expected to occur at the same time with the prompt 
$\gamma$-ray radiation (GRBs). The durations should be  
comparable to these of the associate GRBs. If the Lorentz factors of
shells are distributed between 10 and $10^4$, the typical frequency of
the bulk Compton emission is about 100MeV-GeV. 
Since the primary photons are scattered to
frequencies higher by a boost factor $\propto (\Gamma/\gamma)^2$ where
$\Gamma$ and $\gamma$ are respectively the Lorentz factors of the
scatter and emitter shells, the typical frequency of the bulk Compton
emission for $\gamma_{max}=10^3$ is lower, and the emission peaks around
10-100 MeV.  

\section{Discussion}
We have studied the bulk Compton-scattering in the internal shock model
by using Monte Carlo simulations. We have shown that the radiation drag
does not affect the conversion efficiency in the internal shocks 
even if fireballs have a very large velocity irregularity.
The bulk Compton-scattering is most efficient when an inner engine
produces  shells with comparable energy but very different Lorentz
factors. Such equal energy cases are favored in 
a recent study on GRB temporal profiles (Nakar \& Piran 2002).
We have obtained about a few $\%$ conversion efficiency of the bulk 
Compton-scattering, provided that the shell's Lorentz factor
varies between 10 and $10^4$. This implies that $E_{iso} \sim
10^{50-51}$ ergs might be radiated as 100MeV-GeV $\gamma$-rays. 

The next generation Gamma-Ray Large Area Space Telescope (GLAST),
scheduled for launch in 2007, will provide unprecedented sensitivity 
in the energy range of about $20$ MeV to about 300 GeV. GLAST should be
able to detect the high energy emission produced by the bulk 
Compton-scattering if it is associated with GRBs. GLAST observations will
provide a variable clue to understand the nature of GRB outflows and the
central engines.   

Although the synchrotron self-inverse Compton-scattering also can produce
high energy  $\gamma$-rays, the typical frequencies ($>$ GeV)
could be much higher than that of the bulk Compton emission especially
when fireballs have large velocity fluctuations as we have assumed in
this Letter. Recent studies suggest that fireballs could be highly 
magnetized (Zhang et al. 2003; Kumar \& Panaitescu 2003; Fan et
al. 2002). If this is the case, the self-inverse Compton emission could
be suppressed. However, the bulk Compton-scattering could still produce
high energy $\gamma$-rays as long as fireballs are not dominated by
Poynting-flux. 

This work is supported by the Eberly Research Funds of Penn State,
by the Center for Gravitational Wave Physics, funded under cooperative
agreement by NSF PHY 01-14375, by a NASA Swift Cycle 1 GI program.

\newpage

\newpage
\begin{deluxetable}{cccc}
\tablewidth{0pt}
\tablecaption{Efficiency\label{tab:tab1}}
\tablehead{
\colhead{Mass Distribution} 
& $\gamma_{max}$
& \colhead{Internal Shocks (No Scattering)}
& \colhead{Bulk Compton}
}
\startdata
Equal Energy  & $10^3$ & 8.1$\pm$0.4\ \%  ($8.2\pm 0.4$\  \%)&  $0.7\pm 0.2$\    \%\\
Equal Energy  & $10^4$ &10.5$\pm$0.4\ \%  ($10.9\pm 0.4$\ \%)&  $2.7\pm 0.6$\    \%\\
Equal Mass    & $10^4$ &24.0$\pm$1.4\ \%  ($24.0\pm 1.4$\ \%)&  $(5.8\pm3.7)\times10^{-2}$\  \%\\
Equal Density & $10^4$ &14.5$\pm$1.4\ \%  ($14.5\pm 1.4$\ \%)&  $(1.0\pm0.9)\times10^{-4}$\  \%
\enddata
\end{deluxetable}
\begin{deluxetable}{ccccc}
\tablewidth{0pt}
\tablecaption{The basic types of two shell collisions \label{tab:tab2}}
\tablehead{
\colhead{Mass Distribution} 
& Type I
& Type II
& Type III
& Type IV
}
\startdata
Equal Energy   & 23.7$\pm$1.4\ \%  & 50.4$\pm$2.8\ \%  
               & 4.0$\pm$1.2\ \%   & 21.9$\pm$1.5\ \%\\
Equal Mass     & 32.1$\pm$1.3\ \%  & 24.4$\pm$2.3\ \%
               & 13.3$\pm$2.1\ \%  & 30.2$\pm$1.4\ \%\\
Equal Density  & 33.8$\pm$1.4\ \%  & 16.5$\pm$2.2\ \%
               & 17.9$\pm$2.5\ \%  & 31.9$\pm$1.5\ \%
\enddata
\end{deluxetable}
 \begin{figure}
\plotone{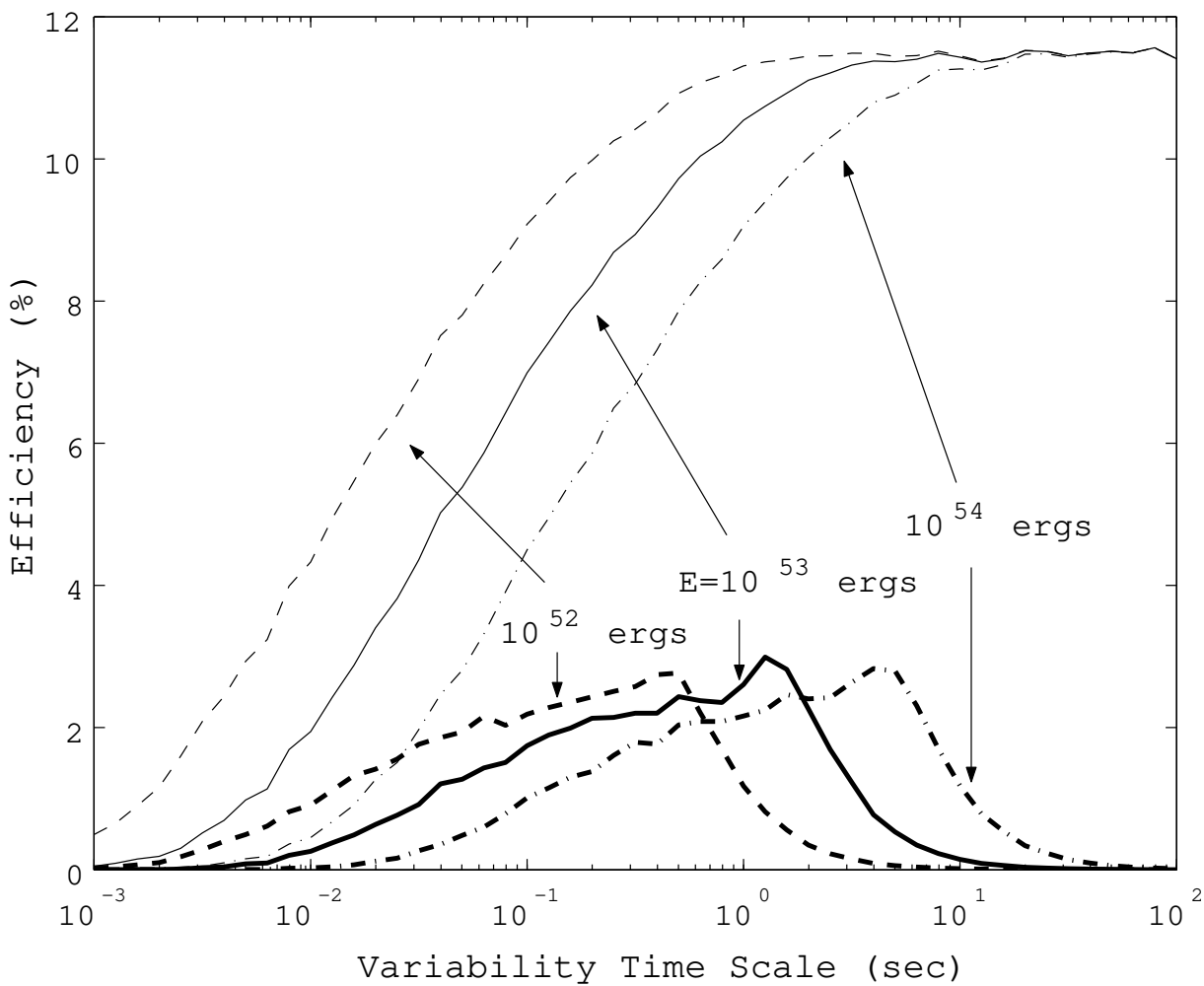}
\caption{Efficiency vs. variability time scale.
Bulk Compton efficiency (thick) and
internal shock efficiency (thin).
$E_{kin,iso}=10^{52}$ ergs (dashed), $10^{53}$ ergs (solid) and 
$10^{54}$ ergs (dashed-dotted).
$300$ equal energy shells, 
$\gamma_{min}=10$ and $\gamma_{max}=10^4$.
\label{fig1}}
 \end{figure}
 \begin{figure}
\plotone{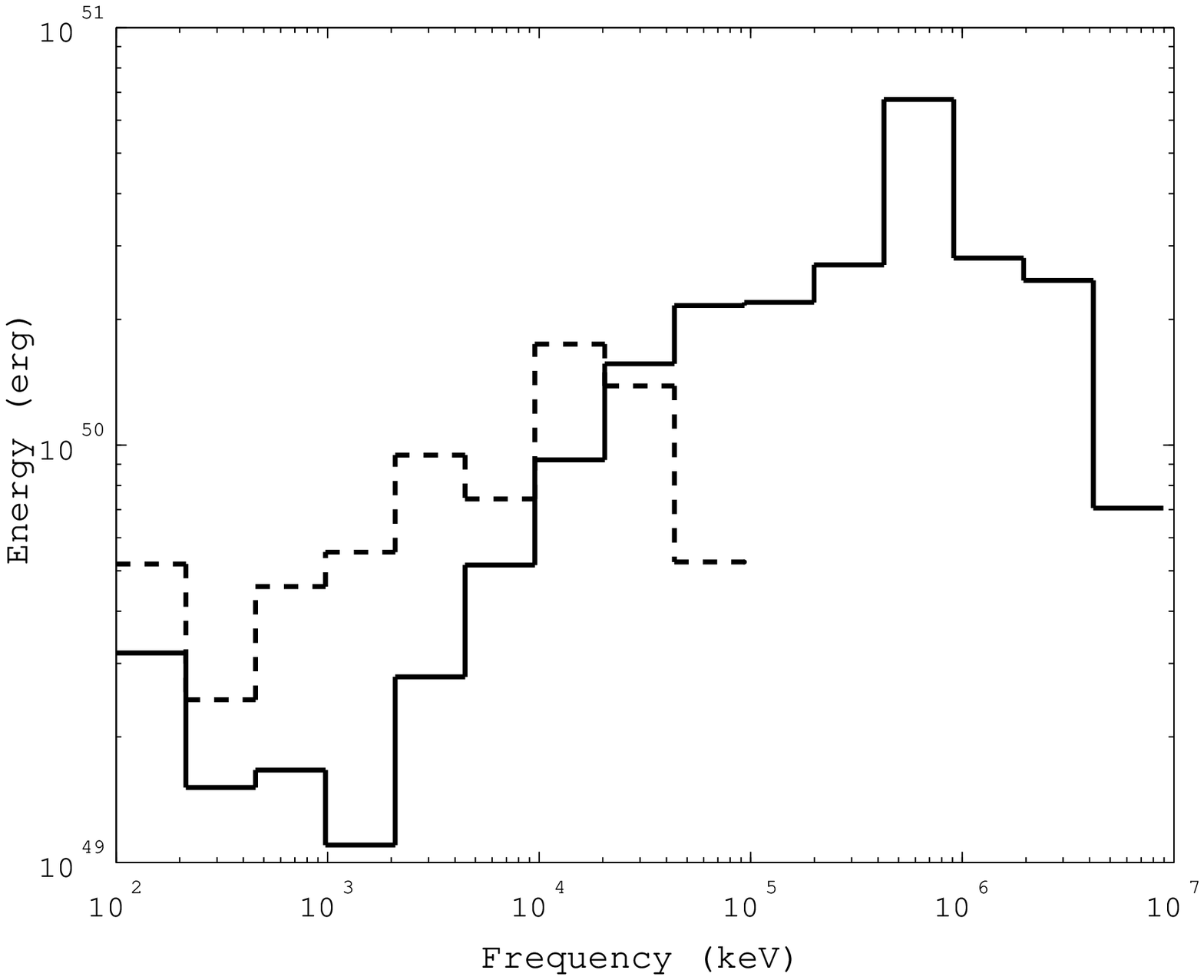}
\caption{Spectra of bulk Compton emission.
$\gamma_{max}=10^3$ (dashed line) or $\gamma_{max}=10^4$ (solid line).
The other parameters are $E_{kin,iso}=10^{53}$ ergs, $300$ equal energy 
shells, $\gamma_{min}=10$ and $l/c=1$ sec.
The cutoff of the solid line 
around a few GeV is due to the Klein-Nishina effect.
\label{fig2}}
 \end{figure}


\begin{references}
\reference{} Beloborodov, A. M.
	2000, ApJ, 539, L25
\reference{} Daigne, F. \& Mochkovitch, R.
	1998, MNRAS, 296, 275
\reference{} Fan,Y.Z.,Dai,Z.G.,Huang,Y.F., \& Lu,T.
	2002, Chinese J.Astron. Astrophys., 2,449
\reference{} Freedman, D. L. \& Waxman, E.
	2001, ApJ, 547, 922
\reference{} Frontera, F. et al.
	2000, ApJS, 127, 59
\reference{} Ghisellini, G.
	2003, "{\it Gamma Ray Bursts in the Afterglow Era - Third Workshop}"
	(astro-ph/0301256)
\reference{} Ghisellini, G.,Lazzati, D., Celotti,A., \& Rees,M.J.
	2000, MNRAS, 316, L45
\reference{} Gruzinov, A. \& M\'{e}sz\'{a}ros, P.
	2000, ApJ, 539, L21
\reference{} Guetta, D., Spada, M. \& Waxman, E.
	2001, ApJ, 557, 399
\reference{} Kobayashi, S., Piran, T. \& Sari, R.
	1997, ApJ, 490, 92
\reference{} Kobayashi, S. \& Sari, R.
	2001, ApJ, 551, 934
\reference{} Kumar, P. \& Panaitescu, A.
	2003, MNRAS, 346, 905
\reference{} Kumar, P. \& Piran, T.
	2000, ApJ, 535, 152
\reference{} Lazzati, D., Ghisellini, G. \& Celotti, A.
	1999, MNRAS, 309, L13
\reference{} Lloyd-Ronning, N. M. \& Zhang, B.
	2004, ApJ, 613, L477
\reference{} Nakar, E. \& Piran, T.
	2002, ApJ, 572, L139
\reference{} Zhang, B., Kobayashi, S. \& \Mesz,P.
	2003, ApJ, 595, 950
\end{references}
\end{document}